\documentclass[11pt]{article}

\usepackage{epsfig}
\usepackage{amssymb}
\usepackage[]{times}
\newcommand{\be}{\begin{equation}}
\newcommand{\ee}{\end{equation}}
\newcommand{\bea}{\begin{eqnarray}}
\newcommand{\eea}{\end{eqnarray}}

\def\1#1{^{(#1)}}

\begin{document}
\title{Rarefaction Shock Waves in Collisionless Plasma \\with Electronic Beam}
\author{Victor Ts. Gurovich${}^{\dag}$ and Leonid G. Fel${}^{\ddag}$\\ \\
${}^{\dag}$Department of Physics, Technion, Haifa 32000, Israel\\
${}^{\ddag}$Department of Civil Engineering, Technion, Haifa 32000, Israel} 
\date{}
\maketitle
\def\be{\begin{equation}}
\def\ee{\end{equation}}
\def\bea{\begin{eqnarray}}
\def\eea{\end{eqnarray}}
\def\p{\prime}
\vspace{-1cm}
\begin{abstract}
We show that an electronic beam passing through the collisionless plasma of the 
"cold" ions and the "hot" Boltzmann electrons can give rise to the propagation 
of the supersonic ion-acoustic rarefaction shock waves. These waves are 
analogous to those predicted by Zeldovich \cite{zr46} in gasodynamics and 
complementary to the ion-acoustic compression shock waves in collisionless 
plasma described by Sagdeev \cite{sa66}.


{\bf PACS 2006:} Primary -- 47.40.-x, Secondary -- 52.35.Tc.
\end{abstract}
\section{Introduction}\label{s1}
It is well known that the supersonic compression shock (CS) waves can propagate
in the ideal gas, but their alternative - the supersonic rarefaction shock (RS)
waves - cannot. In 1937 Zeldovich \cite{zr46} has shown that the interaction 
between the gas molecules possessing the repulsive and attractive parts (e.g., 
the Van der Waals' gas) results in substantial changes: the only RS waves could 
been stably propagated, but the CS waves are unstable. 

In the 1960th Sagdeev \cite{sa66} has predicted the CS wave which may propagate 
in the collisionless (CL) plasma. This appears when the CL plasma comprises the 
"cold" ions and the "hot" Boltzmann electrons and the mean free path is 
determined by the Debye radius of electrons. The shock wave is arisen by the 
supersonic compression soliton disturbed by dissipative effects. Such CS waves 
are analogous to those observed in gasodynamics.

Regarding the RS waves in the CL plasma, a study of such waves dates back to the
papers \cite{be78} and \cite{wa78} for the media comprised the electrons with 
bi-Maxwellian distribution function and "cold" ions. 

In the present article we suggest another kind of the RS waves generation in the
CL plasma based on the supersonic rarefaction solitons. We consider a problem 
in two steps. First, we show that the supersonic rarefaction soliton (of density
and electric potential) can propagate in the CL plasma supplemented by the 
electronic beam. Next, being partially reflected by the soliton's electric 
potential, the beam disturbs its symmetric shape in such a way, that behind the 
maximal value of potential there appear nonlinear oscillations.
%%%%%%%%%%%%%%%%%%%%%%%%%%%%%%%%%%%%%%%%%%%%%%%%%%%
\section{Setup of the Problem}\label{s2}
%%%%%%%%%%%%%%%%%%%%%%%%%%%%%%%%%%%%%%%%%%%%%%%%%%%
Consider the CL plasma with the Boltzmann distribution $N_e=N_{eo}\exp\left(e
\phi(x)/\kappa T_e\right)$ of the "hot" electrons, where $N_{eo}$ is a 
concentration of hot electrons in homogeneous plasma, $T_e$ is a temperature of 
electrons, $e$ and $\kappa$ stand for the charge of electron and the Boltzmann 
constant, respectively. The electric potential $\phi(x)$ is varying along the 
$x$ direction together with the charge density. Such plasma with electronic beam
can be realized in experiments with a hollow anode plasma source \cite{gg06}. 
In what follows, we use the renormalized potential $\psi(x)=-|e|\phi(x)/\kappa 
T_e$ and choose the reference frame related to the electric potential and 
normalized coordinate $\xi=x/D$ where $D=\sqrt{\varepsilon_0 T_e/N_{io}e^2}$ 
denotes the Debay radius and $N_{io}$ and $\varepsilon_0$ stand for homogeneous 
ions density and dielectric vacuum permittivity, respectively. In this reference
frame the ions are running with velocity $V(\xi)$ toward the wave front which is
defined by requirement $V\to U$ when $\psi\to 0$. Here $U$ denotes the velocity
of the homogeneous flow of plasma.

Equations of the energy and mass conservations for cold ions describe their 
1-dim stationary motion and lead \cite{sa66} to the non homogeneous ions density
$N_i(\xi)$,
\bea
N_i(\xi)=\frac{N_{io}}{\sqrt{1+2\psi(\xi)/{\sf M}^2}}\;,\quad {\sf M}=\frac{U}
{C}\;,\quad C=\sqrt{\frac{\kappa T_e}{m_i}}\;,\label{r1}
\eea
where $m_i$ and $C$ denote the mass of the ion and the velocity of the ion 
sound. The Max number for the ions motion is denoted by ${\sf M}$. 

Let ${\cal E}=mv^2/(2\kappa T_e)+\psi(\xi)$ be an energy of the single electron 
of the beam in the $\kappa T_e$ units. Then the density $N_b(\psi)$ of the beam 
electrons is given by the distribution function $f(\xi,v)$,
\bea
N_b(\xi)=\int_0^{\infty}f(\xi,v)dv=\sqrt{\frac{\kappa T_e}{2m}}\int_0^{\infty}
\frac{f({\cal E})d{\cal E}}{\sqrt{{\cal E}-\psi(\xi)}}\;,\quad N_{bo}=\sqrt{
\frac{\kappa T_e}{2m}}\int_0^{\infty}\frac{f({\cal E})d{\cal E}}{\sqrt{{\cal E}
}}\;,\label{r2}
\eea
where $N_{bo}$ denotes a density of the electronic beam in homogeneous plasma, 
and $m$ and $v$ stand for mass and velocity of electron.

The dimensionless potential $\psi(\xi)$ is a smooth function and obeys the 
Poisson equation,
\bea
\frac{d^2\psi}{d\xi^2}=\frac1{\sqrt{1+2\psi(\xi)/{\sf M}^2}}-Ae^{-\psi(\xi)}-
B(\xi)\;,\label{r3}
\eea
where 
\bea
B(\xi)=\frac{N_b(\xi)}{N_{io}}\;,\quad B(0)=\frac{N_{bo}}{N_{io}}\;,\quad 
A=\frac{N_{eo}}{N_{io}}\;,\quad A+B(0)=1\;,\quad A<1\;.\label{r4}
\eea
The last equality holds due to the quasineutrality of homogeneous plasma.

Regarding $N_b(\psi)$, in this paper we deal with two different types of the 
density distribution of electronic beam in accordance with two posed problems: 
the existence of the supersonic rarefaction soliton (see section \ref{s3}) and 
appearance of the supersonic RS waves (see section \ref{s4}).
%%%%%%%%%%%%%%%%%%%%%%%%%%%%%%%%%%%%%%%%%%%%%%%%%%%
\section{The Rarefaction Soliton}\label{s3}
%%%%%%%%%%%%%%%%%%%%%%%%%%%%%%%%%%%%%%%%%%%%%%%%%%%
Consider the electronic beam with a distribution function $f({\cal E})$ given as
follows,
\bea
f({\cal E})>0\quad\mbox{if}\quad\psi\ll{\cal E}_1\leq {\cal E}\leq {\cal E}_2
\quad\mbox{and}\quad f({\cal E})=0\quad\mbox{if}\quad{\cal E}<{\cal E}_1\;
\mbox{or}\;{\cal E}>{\cal E}_2\;.\label{r5}
\eea
Then, by (\ref{r2}), (\ref{r4}) and (\ref{r5}) we have $B(\xi)\simeq B(0)$, and 
equation (\ref{r3}) can been integrated as follows,
\bea
\frac1{2}\left(\frac{d\psi}{d\xi}\right)^2={\sf M}^2\left(\sqrt{1+\frac{2\psi(
\xi)}{{\sf M}^2}}-1\right)+A\left(e^{-\psi}-1\right)-B(0)\psi+K\;,\quad K=const
\;.\label{r6}
\eea
Equation (\ref{r6}) gives rise to soliton solution if $d\psi/d\xi\to 0$ when 
$\psi\to 0$, that immediately requires $K=0$. Rewrite equation (\ref{r6}) for 
$\psi\ll 1$ and preserving the $\psi$- and $\psi^2$ terms, 
\bea
\frac1{2}\left(\frac{d\psi}{d\xi}\right)^2=\psi(1-A+B(0))+\frac{\psi^2}{2}
\left(A-\frac1{{{\sf M}^2}}\right)\;.\label{r7}
\eea
The linear in $\psi$ term in (\ref{r7}) is vanishing due to (\ref{r4}), and what
is left results in ${\sf M}\geq 1/\sqrt{A}$. Keeping in mind $A<1$ we get the 
supersonic rarefaction soliton, ${\sf M}>1$. 

\begin{figure}[h]%[t]
\centerline{\psfig{figure=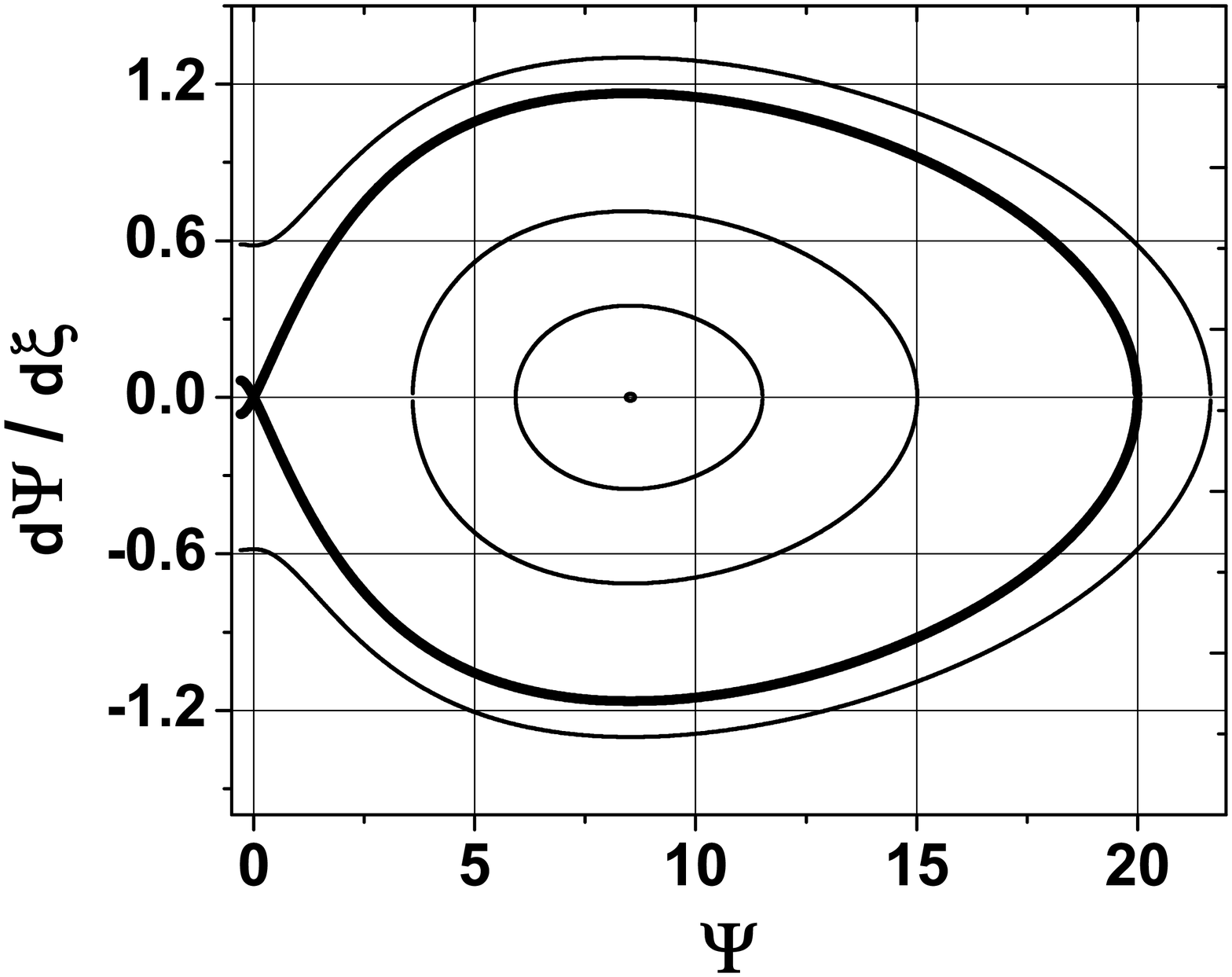,height=5cm,width=10cm}}
\caption{Phase portrait of Eqn (\ref{r7}) for travelling waves of rarefaction; 
a soliton is drawn in bold.}\label{f1}
\end{figure}

Since $\psi(\xi)$ is a smooth function everywhere (also when $\psi\geq 1$) and 
$\psi(\xi)$ arrives its maximal value $\psi_{max}$ then by (\ref{r6}) we get
\bea
{\sf M}^2\left(\sqrt{1+\frac{2\psi_{max}}{{\sf M}^2}}-1\right)+A\left(e^{
-\psi_{max}}-1\right)-B(0)\psi_{max}=0\;.\label{r8}
\eea
The last equation can be resolved analytically as ${\sf M}={\sf M}(A,B(0),
\psi_{max})$. E.g., for the experimental date \cite{gg06} $B(0)\simeq 0.3$, 
$A\simeq 0.7$ and $\psi_{max}\simeq 20$ we obtain ${\sf M}\simeq 1.035$. In 
Figure \ref{f1} we present the phase portrait of equation (\ref{r7}) for the 
parameters given above. The center ({\em the singular point}) is corresponded to
$K\simeq 0.579$ while the separatrix ({\em soliton}) is corresponded to $K=0$.

Up to the sign of the potential $\psi(\xi)$ the phase portrait at Figure 
\ref{f1} is similar to that of the nonlinear ion-acoustic travelling waves of 
compression \cite{sa66}. The main difference appears: in the latter case the 
maximal value of the ion density $N_i(\xi)$ and the maximal value of the 
potential $\psi(\xi)$ occur at the same coordinate $\xi$, while in the former 
case the minimal value of the ion density occur when the potential arrives it 
maximal value. This is why the soliton solution $\psi(\xi)$ is accompanied by 
the ionic rarefaction.
%%%%%%%%%%%%%%%%%%%%%%%%%%%%%%%%%%%%%%%%%%%%%%%%%%%
\section{The Rarefaction Shock Waves}\label{s4}
%%%%%%%%%%%%%%%%%%%%%%%%%%%%%%%%%%%%%%%%%%%%%%%%%%%
In the model of the CS waves \cite{sa66} a symmetric shape of soliton is 
disturbed when a small portion of ions (with a lower energy) is reflected by 
potential barrier, but a large portion of ions (with a higher energy) is passed 
throughout it. In our case the rarefaction soliton will be disturbed by 
electrons of the beam which are reflected by the barrier.

Choose the step-like distribution function $f({\cal E})$,
\bea
f({\cal E})=\sqrt{\frac{8m}{\kappa T_e}}\frac{N_{bo}}{\sqrt{{\cal E}_2}-\sqrt{
{\cal E}_1}}\;\Theta({\cal E}_2-{\cal E})\Theta({\cal E}-{\cal E}_1)\;,
\label{r9}
\eea
where $\Theta(a)$ denotes the Heviside step function. Substituting (\ref{r9}) 
into (\ref{r2}) we get a relative density of the transmitted electronic beam,
\bea
\vartheta_1(\xi)=\frac{B(0)}{2(\sqrt{{\cal E}_2}-\sqrt{{\cal E}_1})}\int_{
{\cal E}_1}^{{\cal E}_2}\frac{d\;{\cal E}}{\sqrt{{\cal E}-\psi(\xi)}}=B(0)\frac{
\sqrt{{\cal E}_2-\psi(\xi)}-\sqrt{{\cal E}_1-\psi(\xi)}}{\sqrt{{\cal E}_2}-
\sqrt{{\cal E}_1}}\;\quad {\cal E}_2>{\cal E}_1\;.\label{r10}
\eea

\begin{figure}[h]%[t]
\centerline{\psfig{figure=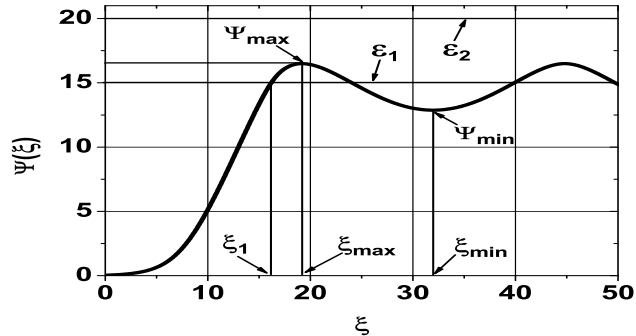,height=5cm,width=10cm}}
\vspace{-.2cm}
\caption{The electric potential $\psi(\xi)$ for the nonlinear RS wave. Its 1st 
maximum $\psi_{max}$ satisfies inequality ${\cal E}_1<\psi_{max}<{\cal E}_2$.}
\label{f2}
\end{figure}

Formula (\ref{r10}) holds for the monotone growing potential $\psi(\xi)$ when 
$\psi(\xi)\leq {\cal E}_1\leq {\cal E}\leq {\cal E}_2$, or $\xi\leq\xi_1$, where
$\psi(\xi_1)={\cal E}_1$ (see Figure \ref{f2}). Here the incident electrons do 
not reflected by barrier but are transmitted if ${\cal E}>{\cal E}_1$.

Next, let us focus on the other case, ${\cal E}_1<\psi(\xi)<{\cal E}<{\cal E}_
2$, where $\psi_{max}=\psi(\xi_{max})$ and $\xi_{max}$ stands for location of 
maximal soliton potential. Here the beam is partly penetrated into the soliton 
potential when $\xi_1\leq\xi\leq\xi_{max}$, but the rest of electrons are 
reflected from the barrier when $\psi(\xi)\geq {\cal E}_1$, or $\xi\geq\xi_1$. 
It leads to the changes in formula (\ref{r10}) 
\bea
\vartheta_2(\xi)={\overline B}\sqrt{{\cal E}_2-\psi(\xi)}\;,\quad\xi_1\leq
\xi\leq\xi_{max}\;,\quad {\overline B}=\frac{B(0)}{\sqrt{{\cal E}_2}-\sqrt{{
\cal E}_1}}\;,\label{r11}
\eea

Finally, consider the last case $\psi_{max}<{\cal E}<{\cal E}_2$ (electrons do 
not reflect by the barrier)
\bea
\vartheta_3(\xi)={\overline B}\left(\sqrt{{\cal E}_2-\psi(\xi)}-\sqrt{\psi_{max}
-\psi(\xi)}\right)\;.\label{r12}
\eea

For convenience, unify three functions $\vartheta_i(\xi)$, $i=1,2,3$, given in 
different ranges of $\xi$ by one $I(\xi)$ given in whole range of $\xi$,
\bea
I(\xi)&=&{\overline B}\left(\sqrt{{\cal E}_2-\psi(\xi)}-\Theta\left({\cal 
E}_1-\psi(\xi)\right)\Theta\left(\xi_{max}-\xi\right)\sqrt{{\cal E}_1-\psi(\xi)
}-\right.\nonumber\\
&&\hspace{5cm}\left.\Theta\left(\xi-\xi_{max}\right)\sqrt{\psi_{max}-\psi(\xi)}
\right).\label{r13}
\eea
In the similar way we can construct the relative density $R(\xi)$ of the 
electronic beam reflected from the barrier,
\bea
R(\xi)={\overline B}\left(\sqrt{\psi_{max}-\psi(\xi)}-\Theta\left({\cal E}_1-
\psi(\xi)\right)\sqrt{{\cal E}_1-\psi(\xi)}\right)\;,\quad\xi\leq\xi_{max}\;.
\label{r14}
\eea

A choice of arguments in the $\Theta$-functions in (\ref{r13}) and (\ref{r14})
is motivated by need to represent the Poisson equation (\ref{r3}) as an 
autonomous differential equation.

Substitute the entire density $B(\xi)=I(\xi)+R(\xi)$ of the beam into equation 
(\ref{r3}) and obtain its 1st integral,
\bea
\frac1{2}\left(\frac{d\psi}{d\xi}\right)^2-V_{-}\left(\psi\right)=0\;,\quad
\xi\leq\xi_{max}\;,\label{r15}
\eea
where the quasipotential $V_{-}(\psi)$ is given by
\bea
V_{-}(\psi)&=&{\sf M}^2\left(\sqrt{1+\frac{2\psi}{{\sf M}^2}}-1\right)+A\left(
e^{-\psi}-1\right)+\label{r16}\\
&&\frac{2{\overline B}}{3}\left[({\cal E}_2-\psi)^{3/2}+(\psi_{max}-\psi)^{
3/2}-2({\cal E}_1-\psi)^{3/2}\Theta({\cal E}_1-\psi)+2{\cal E}_1^{3/2}-
{\cal E}_2^{3/2}-\psi_{max}^{3/2}\right].\nonumber
\eea
In (\ref{r15}) we have taken a zero's value for the integration constant (see 
(\ref{r6}) with $K=0$) to provide the requirement $d\psi/d\xi\to 0$ when 
$\psi\to 0$. 

Rewrite the quasipotential (\ref{r16}) for $\psi\ll 1$ and preserve the 
non-linear terms up to $\psi^2$,
\bea
V_{-}(\psi)&=&\psi\left[1-A-{\overline B}\left(\sqrt{{\cal E}_2}+\sqrt{\psi_{
max}}-2\sqrt{{\cal E}_1}\right)\right]\label{r17}\\
&+&\frac{\psi^2}{2}\left[-\frac1{{\sf M}^2}+A+\frac{{\overline B}}{2}\left(
\frac1{\sqrt{{\cal E}_2}}+\frac1{\sqrt{\psi_{max}}}-\frac{2}{\sqrt{{\cal E}_1}}
\right)\right]\;.\nonumber
\eea
Note that by (\ref{r10},\ref{r14}) the term ${\overline B}\left(\sqrt{{\cal E}_
2}+\sqrt{\psi_{max}}-2\sqrt{{\cal E}_1}\right)={\overline B}\left(\sqrt{{\cal 
E}_2}-\sqrt{{\cal E}_1}\right)+{\overline B}\left(\sqrt{\psi_{max}}-\sqrt{{\cal 
E}_1}\right)$ which enters into the linear in $\psi$ term in (\ref{r17}), gives 
a total density of the electronic beam including incidence and reflection as 
well. Then the whole linear in $\psi$ term in (\ref{r17}) disappears due to the 
quasineutrality of homogeneous plasma in the general case (when beam's 
reflection exists),
\bea
A+{\overline B}\left(\sqrt{{\cal E}_2}+\sqrt{\psi_{max}}-2\sqrt{{\cal E}_1}
\right)=1\quad\rightarrow\quad A<1\;.\label{r18}
\eea
Substituting the quadratic in $\psi$ term of (\ref{r17}) into (\ref{r15}) we 
find the lower bound for ${\sf M}$,
\bea
\frac1{{\sf M}^2}\leq A+\frac{{\overline B}}{2}\left(\frac1{\sqrt{{\cal E}_2
}}+\frac1{\sqrt{\psi_{max}}}-\frac{2}{\sqrt{{\cal E}_1}}\right)\quad\rightarrow
\quad\frac1{{\sf M}^2}<A\;.\label{r19}
\eea
Combining (\ref{r18}) and (\ref{r19}) we arrive at ${\sf M}>1/A>1$, i.e., the 
supersonic RS wave.

An exact value of the Mach number can be found if we consider equation 
(\ref{r15}) for $\xi=\xi_{max}$ and $\psi=\psi_{max}>{\cal E}_1$, i.e., when 
$d\psi/d\xi=0$,
\bea
\sqrt{1+\frac{2\psi_{max}}{{\sf M}^2}}-1=\frac{A}{{\sf M}^2}\left(1-e^{-
\psi_{max}}\right)+\frac{2{\overline B}}{3{\sf M}^2}\left[{\cal E}_2^{3/2}+
\psi_{max}^{3/2}-({\cal E}_2-\psi_{max})^{3/2}-2{\cal E}_1^{3/2}\right]\;.
\quad\label{r20}
\eea
The last equality together with a quasineutrality condition (\ref{r18}) allow to
obtain ${\sf M}$ and $A$ if the other four parameters ${\overline B}$, 
${\cal E}_1$, ${\cal E}_2$ and $\psi_{max}$ are given.

Now, consider the Poisson equation (\ref{r3}) in the range $\xi\geq \xi_{max}$
\bea
\frac1{2}\left(\frac{d\psi}{d\xi}\right)^2-V_{+}\left(\psi\right)=0\;,\quad
\xi\geq\xi_{max}\;,\label{r21}
\eea
where the quasipotential $V_{+}(\psi)$ is given by
\bea
V_{+}(\psi)&=&{\sf M}^2\left(\sqrt{1+\frac{2\psi}{{\sf M}^2}}-\sqrt{1+\frac{2
\psi_{max}}{{\sf M}^2}}\right)+A\left(e^{-\psi}-e^{-\psi_{max}}\right)+
\label{r22}\\
&&\frac{2{\overline B}}{3}\left[({\cal E}_2-\psi)^{3/2}-(\psi_{max}-\psi)^{3/2}
-({\cal E}_2-\psi_{max})^{3/2}\right]\;.\nonumber
\eea
In (\ref{r22}) we have taken non-zero's value for the integration constant (see 
(\ref{r6}) with $K\neq 0$) to provide the requirement $V_{+}(\psi_{max})=0$. 
Note that if the potential $\psi(\xi)$ arrives also its minimal value $\psi_{
min}$ at $\xi_{min}>\xi_{max}$ (see Figure \ref{f2}), then $V_{+}(\psi_{min})=0$
as it follows from (\ref{r21}). By equalities (\ref{r18}), (\ref{r20}) and 
(\ref{r22}) it follows also that the value $\psi_{min}$ is completely 
determined by four parameters ${\overline B}$, ${\cal E}_1$, ${\cal E}_2$ and 
$\psi_{max}$.
\begin{figure}[h]%[t]
\centerline{\psfig{figure=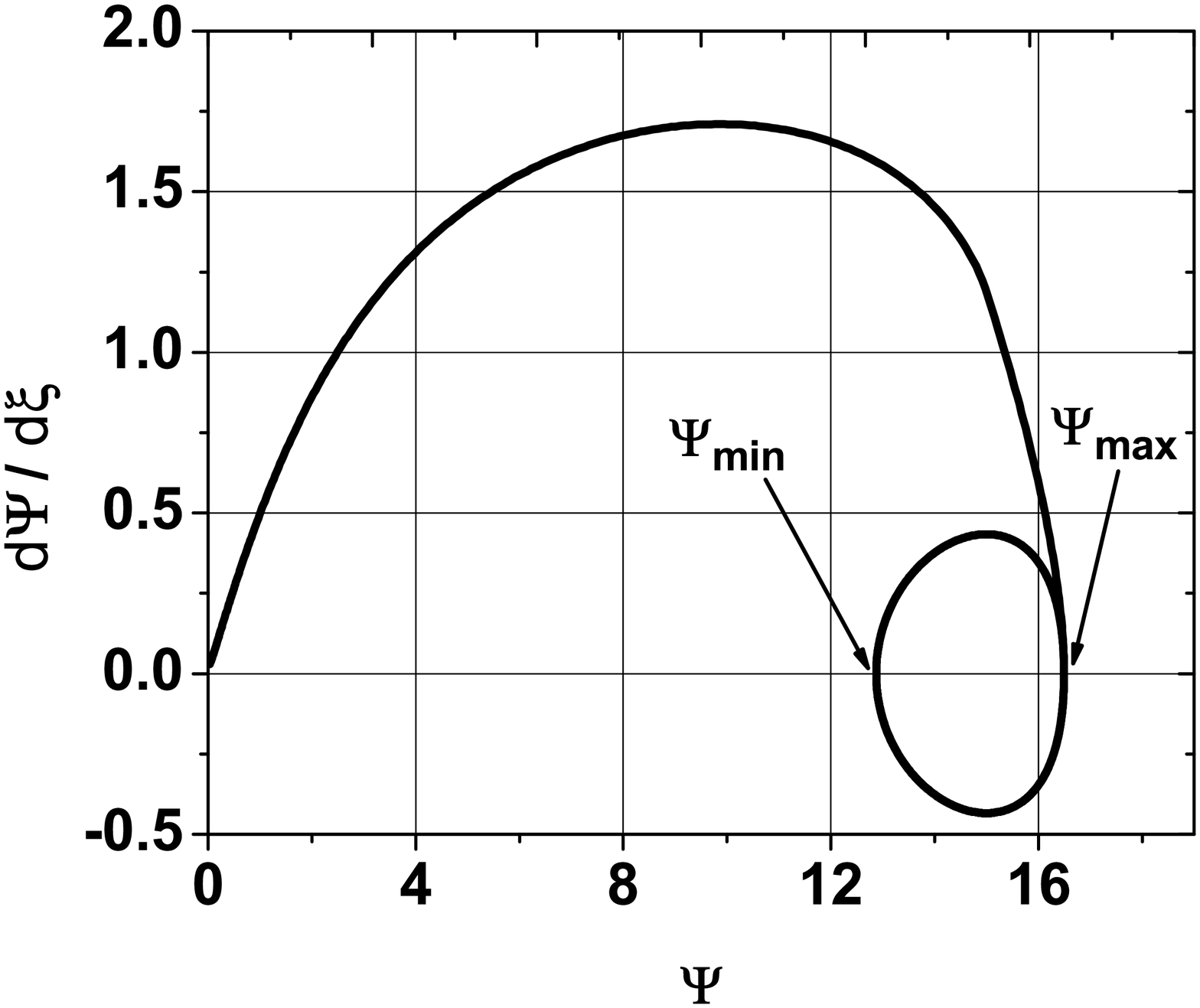,height=5.5cm,width=10cm}}\vspace{-.2cm}
\caption{Phase portrait of equations (\ref{r15}) and (\ref{r21}) for the RS
waves.}\label{f3}
\end{figure}

In Figure \ref{f3} we present the phase portraits of two first integrals 
(\ref{r15}) and (\ref{r21}) of the Poisson equation (\ref{r3}) for the date
${\cal E}_1=15$, ${\cal E}_2=20$, $\psi_{max}\simeq 16.5$ and $B(0)\simeq 0.15$
taken from experiments \cite{gg06}. The values $A\simeq 0.803$, ${\sf M}\simeq 
1.433$ and $\psi_{min}\simeq 12.86$ were found by (\ref{r18}), (\ref{r20}) and 
(\ref{r22}), respectively.

Both portraits are unified in one smooth curve. Its 1st part (an arch in the 
upper half-plane, $d\psi/d\xi\geq 0$) is related to the soliton-like behavior of
the electric potential $\psi$ at the front of the RS waves, while its 2nd part 
(a closed loop) is related to the nonlinear oscillations of $\psi$.
\begin{figure}[h]%[t]
\centerline{\psfig{figure=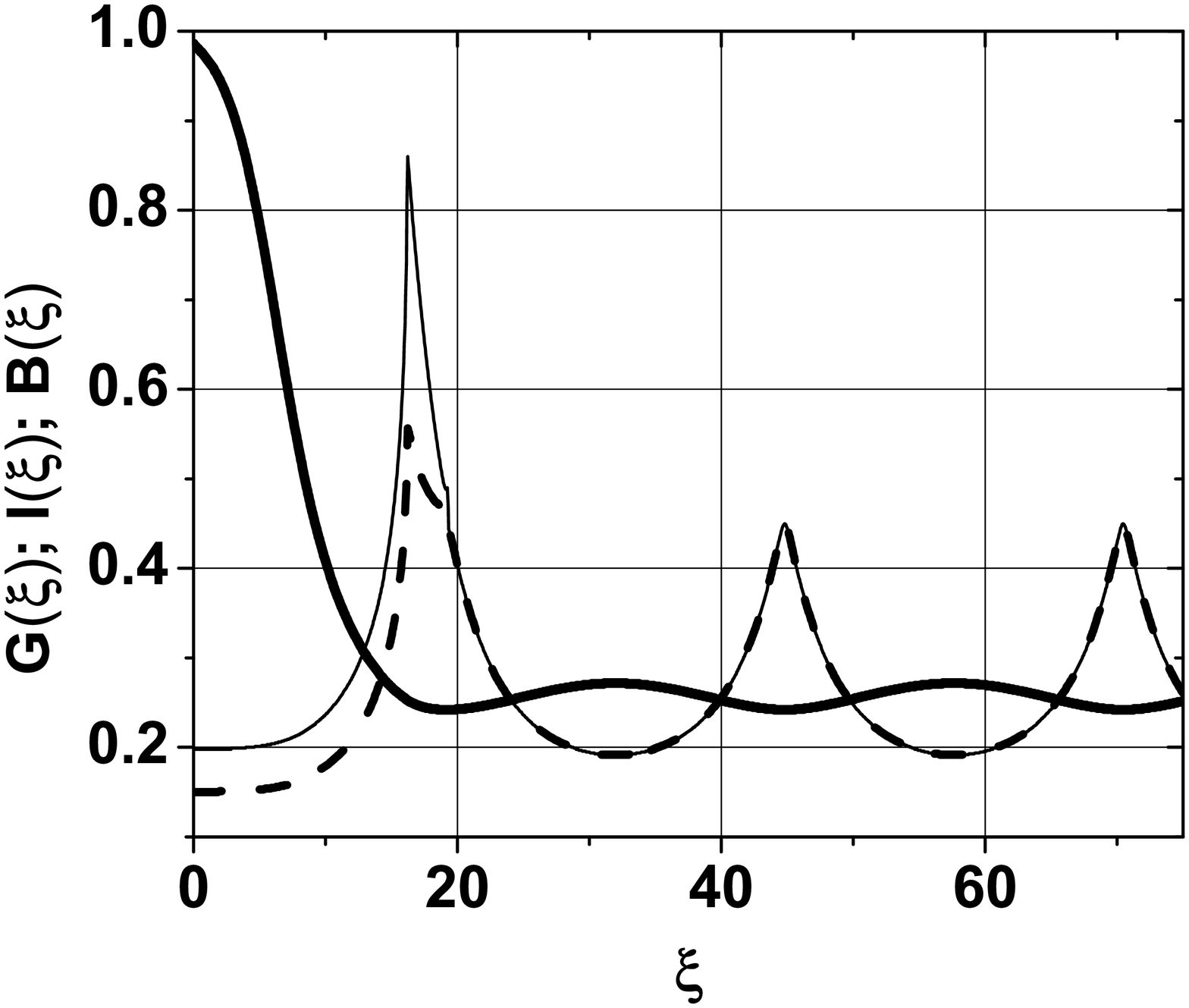,height=5cm,width=10cm}}\vspace{-.2cm}
\caption{Densities' distribution of the total electronic beam $B(\xi)$,
including reflections ({\em plain}); of the transmitted electronic beam $I(
\xi)$ ({\em dashed}), and of the ions $G(\xi)$ ({\em bold}).}\label{f4}
\end{figure}

In Figure \ref{f4} we present three densities' distributions: the total density
of the electronic beam $B(\xi)$ when the reflected electrons are accounted for;
the density of the transmitted electronic beam $I(\xi)$ and also the density of 
the ions in plasma, $G(\xi)$. The maximal value of $G(\xi)$ is arrived before 
the wave front ($\xi=0$) of electric potential $\psi(\xi)$, while the density of
the ions in the region $\xi>0$ is much less. This manifests that we get the 
rarefaction shock wave.

We appreciate useful discussions with L.P.Pitaevsky, Y.E.Krasik and 
I.D.Kaganovich.

\end{document}